\documentclass[a4paper,12pt]{article}
\usepackage{amsmath}
\usepackage{amssymb}
\usepackage{amsfonts}
\usepackage{fleqn}
\usepackage[footnotesize]{caption}

\def\beq{\begin{equation}}
\def\eeq{\end{equation}}
\def\bea{\begin{eqnarray}}
\def\eea{\end{eqnarray}}

\def\<{\left\langle}
\def\>{\right\rangle}

\allowdisplaybreaks[2]
\begin{document}

\begin{titlepage}

 \vspace*{-15mm}
\begin{flushright}
\end{flushright}
\vspace*{5mm}

\begin{center}
{ \sffamily \LARGE Tri-bimaximal neutrino mixing from discrete
subgroups of $SU(3)$ and $SO(3)$ family symmetry}
\\[8mm]

I.~de~Medeiros~Varzielas
\footnote{E-mail:\texttt{i.varzielas@physics.ox.ac.uk}}$^{(a)}$,
S.~F.~King
\footnote{E-mail:\texttt{sfk@hep.phys.soton.ac.uk}}$^{(b)}$
and G.~G.~Ross
\footnote{E-mail:\texttt{g.ross@physics.ox.ac.uk}}$^{(a)}$
\\[3mm]

{\small\it
$^{(a)}$
Department of Physics, Theoretical Physics, University of Oxford\\
1 Keble Road, Oxford OX1 3NP, U.K.}
\\[1mm]
{\small\it
$^{(b)}$ School of Physics and Astronomy,
University of Southampton,\\
Southampton, SO17 1BJ, U.K.}
\\[1mm]
\end{center}
\vspace*{0.75cm}

\begin{abstract}
\noindent It has recently been shown how tri-bimaximal neutrino
mixing can be achieved, using the see-saw mechanism with
constrained sequential dominance, through the vacuum alignment of
a broken non-Abelian gauged family symmetry such as $SO(3)$ or
$SU(3)$. Generalising the approach of Altarelli and Feruglio
developed for an $A_4$ model we show how the reduction of the
underlying symmetry to a discrete subgroup of $SO(3)$ or $SU(3)$
renders this alignment a generic property of such models. This
means near tri-bimaximal mixing can be quite naturally
accommodated in a complete unified theory of quark and lepton
masses.
\end{abstract}

\end{titlepage}\newpage \setcounter{footnote}{0}

\section{Introduction}

Current neutrino oscillation results \cite{Maltoni:2004ei} are
consistent with so-called tri-bimaximal lepton mixing 
in which the lepton mixing matrix takes the approximate Harrison,
Perkins, Scott \cite{tribi} form:
\begin{equation}
U_{HPS}\approx
\begin{pmatrix}
\sqrt{\frac{2}{3}} & \frac{1}{\sqrt{3}} & 0 \\
-\frac{1}{\sqrt{6}} & \frac{1}{\sqrt{3}} & \frac{1}{\sqrt{2}} \\
\frac{1}{\sqrt{6}} & -\frac{1}{\sqrt{3}} & \frac{1}{\sqrt{2}}%
\end{pmatrix}%
.  \label{tribi}
\end{equation}%
Given the uncertainties in the current measured values of the neutrino
mixing angles, and the theoretical corrections inherent in any model of
lepton mixing, it is likely that tri-bimaximal mixing, if at all relevant,
is realised only approximately. Nevertheless, given the symmetrical nature
of tri-bimaximal mixing, it is of interest to see if it can be reproduced,
at least approximately, in models of quark and lepton masses and mixings, in
particular those based on the see-saw mechanism where the smallness of
neutrino masses emerges most elegantly.

The fact that the MNS mixing matrix in Eq.(\ref{tribi}) involves
square roots of simple ratios motivates models in which the mixing
angles are independent of the mass eigenvalues. One such class of
models are see-saw models with sequential dominance (SD) of
right-handed neutrinos \cite{King:1998jw}. In SD, the atmospheric
and solar neutrino mixing angles are determined in terms of ratios
of Yukawa couplings involving those right-handed neutrinos which
give the dominant and subdominant contributions, respectively, to
the see-saw mechanism. If the Yukawa couplings involving different
families are related in some way, then it is possible for neutrino
mixing angle relations to emerge in a simple way, independently of
the neutrino mass eigenvalues. For example, maximal atmospheric
neutrino mixing results from the Yukawa couplings involving second
and third families having equal Yukawa couplings (up to a phase)
to the dominant right-handed neutrino. Tri-bimaximal neutrino
mixing then follows if, in addition, the Yukawa couplings
involving all three families couple democratically to the leading
subdominant right-handed neutrino, providing the couplings are
relatively real and the second or third coupling is in anti-phase
relative to those of the dominant couplings
\cite{ggrparis,King:2005bj,deMedeirosVarzielas:2005ax}. If the
dominant and subdominant right-handed neutrinos dominate the
see-saw mechanism by virtue of their lightness, then they may have
the smallest Yukawa couplings, and such democratic relations
between different families would not be readily apparent in the
charged fermion Yukawa matrices.

The above picture in which Yukawa couplings of different families
are equal (up to phases) strongly suggests a non-Abelian family
symmetry which is acting behind the scenes to relate all three
families together, as emphasised in \cite{ggrparis,GG}. In the
charged fermion sector, the presence of such a non-Abelian family
symmetry is well hidden from view since the masses of the three
families of charged fermions are strongly hierarchical, and thus
any non-Abelian family symmetry must be strongly and
hierarchically broken. Even though the family symmetry is strongly
broken, it is possible for the required equalities of Yukawa
couplings to emerge if the several scalar fields which break the
family symmetry (called flavons) have their vacuum expectation
values (VEVs) carefully (mis)aligned along special directions in
family space. Then, if these flavons appear in the effective
operators responsible for the Yukawa couplings, the equality of
the Yukawa couplings in the SD picture may be due to the
particular vacuum alignment of the flavons responsible for that
particular operator. These ingredients have been recently used as
the basis for models of quark and lepton masses and mixings,
incorporating tri-bimaximal neutrino mixing,
based on $SO(3)$ \cite{King:2005bj} and $SU(3)$ \cite%
{deMedeirosVarzielas:2005ax} family symmetry. However it must be admitted
that in these models the vacuum alignment is not realised in the most
elegant or efficient manner, and it is one of the purposes of this note to
show that the physics of vacuum alignment simplifies if the continuous
family symmetry is replaced by a discrete family symmetry subgroup.

In this letter, then, we discuss how the vacuum (mis)alignment needed for
tri-bimaximal mixing proceeds quite readily in the case that the theory is
invariant under a discrete subgroup of either $SO(3)$ or $SU(3)$ family
symmetry. Our vacuum alignment mechanism is a related to that of Altarelli
and Feruglio who analysed the spontaneous breaking of $A_{4}$ \cite%
{Altarelli:2005yp}, and indeed we show that it immediately allows for a
4-dimensional version of the $A_{4}$ model\footnote{%
This has been noted by Altarelli and Feruglio\cite{altlatest} in a
recent paper that was issued during the completion of this paper.}
without supernatural fine tuning. However our main focus is
concerned with
simplifying the $SO(3)$ and $SU(3)$ models of refs \cite%
{King:2005bj,deMedeirosVarzielas:2005ax}. An important distinction between
these models is whether they allow the quadratic invariant $\Sigma _{i}\phi
_{i}\phi _{i}$, as is the case for $SO(3)$ \cite{King:2005bj} or $A_{4}$
\cite{Altarelli:2005yp}, or forbid it as is the case for $SU(3)$ \cite%
{deMedeirosVarzielas:2005ax}. The reason that this is important is that, in
the former case, viable models of fermion mass require that the left-handed $%
SU(2)_{L}$ doublet fermions, $\psi _{i},$ transform differently from the
left-handed charge conjugate $SU(2)_{L}$ singlet fermions, $\psi _{i}^{c}$.
As a result it is not straightforward to implement an underlying $%
SO(10)\otimes G_{family}$ symmetry. If, as is the case for $SU(3)$, the
bilinear invariant is absent then it is possible to achieve this unification
\cite{deMedeirosVarzielas:2005ax}. We present three examples, two which
apply to the \textquotedblleft $SO(3)-$like\textquotedblright case
(including $A_{4}$) and one which applies to the \textquotedblleft $SU(3)-$%
like\textquotedblright case.

\section{Constrained sequential dominance}

\label{Sec:CSD}

To see how tri-bimaximal neutrino mixing could emerge from SD, we begin by
writing the right-handed neutrino Majorana mass matrix $M_{\mathrm{RR}}$ in
a diagonal basis as
\begin{equation*}
M_{\mathrm{RR}}=%
\begin{pmatrix}
X & 0 & 0 \\
0 & Y & 0 \\
0 & 0 & Z%
\end{pmatrix}%
,
\end{equation*}%
where we shall assume
\begin{equation}
X\lesssim Y\ll Z.  \label{LSD}
\end{equation}%
In this basis we write the neutrino (Dirac) Yukawa matrix $Y_{LR}^{\nu }$ in
terms of $(1,3)$ column vectors $A_i,$ $B_i,$ $C_i$ as
\begin{equation}
Y_{\mathrm{LR}}^{\nu }=%
\begin{pmatrix}
A & B & C%
\end{pmatrix}%
  \label{Yukawa}
\end{equation}%
in the convention where the Yukawa matrix corresponds to the Lagrangian
coupling $\bar{L}H_{u}Y_{LR}^{\nu }\nu _{R}$, where $L$ are the left-handed
lepton doublets, $H_{u}$ is the Higgs doublet coupling to up-type quarks and
neutrinos, and $\nu _{R}$ are the right-handed neutrinos. The Dirac neutrino
mass matrix is then given by $m_{\mathrm{LR}}^{\nu }=Y_{LR}^{\nu }v_{\mathrm{%
u}}$, where $v_{\mathrm{u}}$ is the vacuum expectation value (VEV) of $H_{u}$%
. The effective Lagrangian resulting from integrating out the massive right
handed neutrinos is
\begin{equation}
L_{eff}=\frac{(\nu_{i}^{T} A_{i})(A^{T}_{j} \nu_{j})}{X}+\frac{(\nu_{i}^{T} B_{i})(B^{T}_{j} \nu_{j})}{Y}%
+\frac{(\nu_{i}^{T} C_{i})(C^{T}_{j} \nu_{j})}{Z}  \label{leff}
\end{equation}%
where $\nu _{i},i=1,2,3$ are the left handed neutrino fields.

The case of interest here is the one in which these terms are
ordered, due to the ordering in Eq.(\ref{LSD}), with the third
term negligible, the second term subdominant and the first term
dominant - \textquotedblleft light sequential
dominance\textquotedblright\ (LSD)\cite{King:1998jw}, ``light''
because the lightest right-handed neutrino makes the dominant
contribution to the see-saw mechanism. LSD is motivated by unified
models in which only small mixing angles are present in the Yukawa
sector, and implies that the heaviest right-handed neutrino of
mass $Z$ is irrelevant for both leptogenesis and neutrino
oscillations (for a discussion of all these points see
\cite{King:2003jb}).

In \cite{ggrparis,King:2005bj} we proposed the following set of
conditions which are sufficient to achieve tri-bimaximal mixing
within the framework of sequential dominance ``constrained
sequential dominance (CSD)'':
\begin{subequations}
\label{tribiconds}
\begin{eqnarray}
|A_{1}| &=&0,  \label{tribicondsd} \\
\text{\ }|A_{2}| &=&|A_{3}|,  \label{tribicondse} \\
|B_{1}| &=&|B_{2}|=|B_{3}|,  \label{tribicondsa} \\
A^{\dagger }B &=&0.  \label{zero}
\end{eqnarray}

The condition in Eqs.(\ref{tribicondsd},\ref{tribicondse}) gives
rise to bi-maximal mixing in the atmospheric neutrino sector,
$\tan \theta _{23}^{\nu }=1$. The remaining conditions in
Eq.\ref{tribiconds} give
tri-maximal mixing in the solar neutrino sector, $\tan \theta _{12}^{\nu }=1/%
\sqrt{3}$ and to $\theta _{13}^{\nu }=0$.

With this it is straightforward to build theories which generate
tri-bimaximal mixing. A very simple example example is provided by
a
supersymmetric theory in which the lepton doublets $L$ are triplets of an $%
SO(3)$ family symmetry, but the CP conjugates of the right-handed
neutrinos, $\nu _{i}^{c},$ and Higgs doublets, $H_{u,d},$ are
singlets under the family symmetry\footnote{$SO(3)$ has been
previously used as a family symmetry in
e.g.\cite{Barbieri:1999km}.} \cite{Antusch:2004xd}. To generate
hierarchical charged lepton masses we need spontaneous breaking of
the family symmetry
\end{subequations}
\begin{equation}
SO(3)\longrightarrow SO(2)\longrightarrow \mathrm{Nothing}.  \label{fsb}
\end{equation}%
To achieve this symmetry breaking we introduce the additional $SO(3)$
triplet \textquotedblleft flavon\textquotedblright\ fields $%
\phi _{3}$, $\phi _{23},\phi _{123}$ whose VEVs, $<\phi >$ break the $SO(3)$
family symmetry. The vacuum alignment of the flavon VEVs plays a crucial
role in this model, as follows. Suppose that symmetries of the model allow
only theYukawa couplings associated with the superpotential terms of the
form:
\begin{equation}
y^{\prime \prime }LH_{u}\nu _{1}^{c}\frac{\phi _{23}}{M}+yLH_{u}\nu _{2}^{c}%
\frac{\phi _{123}}{M}+y^{\prime }LH_{u}\nu _{3}^{c}\frac{\phi _{3}}{M}
\label{Yuk}
\end{equation}%
where $y,y^{\prime },y^{\prime \prime }$ are complex Yukawa
couplings, $M$ is a mass scale. These generate
Dirac neutrino mass terms of the form given in Eq.(\ref{Yukawa}) with%
\begin{equation*}
A_{i}=<{\phi _{23}>}_{i},\text{ }B_{i}=<{\phi _{123}>}_{i},\text{ }C_{i}=<{%
\phi _{3}>}_{i}.
\end{equation*}%
Provided the vacuum alignment of the VEVs of $\phi _{3}$, $\phi
_{23},\phi
_{123}$ satisfy Eqs.(\ref{tribicondsd}, \ref{tribicondse}, \ref{tribicondsa},%
\ref{zero}) one achieves tri-bimaximal mixing with sequential
dominance.

This example clearly illustrates the importance of this pattern of
vacuum (mis)alignment of the flavon VEVs in achieving
tri-bimaximal mixing and the remainder of this paper is concerned
with achieving such a vacuum (mis)alignment using discrete family
symmetries.

\section{$A_{4}$}

We start with a discussion of the vacuum structure for the potential of a
model of fermion masses based on the discrete symmetry $A_{4}$ $\cite%
{Altarelli:2005yp}.$The group $A_{4}$ (or $\Delta (12)$)\footnote{
$A_{4}\equiv \Delta (12)$ is one of the family of dihedral like
$\Delta (3n^{2})$ finite subgroups of $SU(3)$, whose irreducible
representations are either 1 or 3 dimensional \cite{Klink}.} is a
discrete subgroup of $SO(3)$ and $SU(3)$ and so it is relevant to
the generalisation of the $SO(3)$ and $SU(3) $ family symmetry
models. In this model the symmetry breaking is generated
by two $A_{4}$ triplet fields $\varphi $ and $\varphi ^{\prime }$ with VEVs $%
\varphi =(v,v,v)$ and $\varphi ^{\prime }=(0,0,v^{\prime })$. Although the
notation is different, these correspond to the flavons $\phi _{123}$ and $%
\phi _{3}$ discussed earlier, and this alignment leads to a model of
tri-bimaximal mixing \cite{Altarelli:2005yp}. The alignment is naturally
generated along the $F-$flat direction in a specific model with the
superpotential constrained by an additional $Z_{3}\otimes U(1)_{R}$ symmetry
under which the fields transform as in Table \ref{x}. In addition the model
uses the triplet \textquotedblleft driving\textquotedblright\ fields $%
\varphi _{0}$, $\varphi _{0}^{\prime }$, as well as two\footnote{%
The original model uses just one, c.f. \cite{Altarelli:2005yp}} $A_{4}$
singlets $\xi _{1}$, $\xi _{2}$ that acquire VEVs and $\xi _{0}$ to drive
these. Their charge assignments under $Z_{3}\otimes U(1)_{R}$ are listed in
Table \ref{x}, where $\omega $ is the cube root of unity.

The most general renormalisable superpotential allowed by these symmetries
is given by
\begin{eqnarray}
w_{d} &=&M(\varphi _{0}\varphi )+g(\varphi _{0}\varphi \varphi
)+g_{1}(\varphi _{0}^{\prime }\varphi ^{\prime }\varphi ^{\prime })  \notag
\\
&&+(f_{1}\xi _{1}+f_{2}\xi _{2})\varphi _{0}^{\prime }\varphi ^{\prime
}++f_{3}\xi _{0}(\varphi ^{\prime }\varphi ^{\prime })+f_{ij}\xi _{0}\xi
_{i}\xi {j}  \label{wd}
\end{eqnarray}%
where the 3-triplet invariant $\phi \phi \phi $ stands for $\phi _{1}\phi
_{2}\phi _{3}$ and cyclic permutations.

The vacuum minimisation conditions correspond to the vanishing of the $F-$%
terms. For the $\varphi $ field this corresponds to
\begin{eqnarray}
\frac{\partial w}{\partial \varphi _{01}} &=&M\varphi _{1}+g\varphi
_{2}\varphi _{3}=0  \notag \\
\frac{\partial w}{\partial \varphi _{02}} &=&M\varphi _{2}+g\varphi
_{3}\varphi _{1}=0  \notag \\
\frac{\partial w}{\partial \varphi _{03}} &=&M\varphi _{3}+g\varphi
_{1}\varphi _{2}=0
\end{eqnarray}

\begin{table}[tbp] \centering%
\begin{tabular}{|l||l|l|l|l||l|l|l|}
\hline
\texttt{Field} & $\varphi $ & $\varphi ^{\prime }$ & $\xi _{1}$ & $\xi _{2}$
& $\varphi _{0}$ & $\varphi _{0}^{\prime }$ & $\xi _{0}$ \\ \hline\hline
$Z_{3}$ & $1$ & $\omega $ & $\omega $ & $\omega $ & $1$ & $\omega $ & $%
\omega $ \\
$U(1)_{R}$ & $0$ & $0$ & $0$ & $0$ & $2$ & $2$ & $2$ \\ \hline
\end{tabular}
\caption{Transformation property of the fields in the $A_4$ model.}\label{x}%
\end{table}%

These are solved by
\begin{equation}
\varphi =(v,v,v),~~~~~~~v=-\frac{M}{g}.  \label{phivev}
\end{equation}%
For the $\varphi ^{\prime }$ field the minimisation conditions are given by
\begin{eqnarray}
\frac{\partial w}{\partial \varphi _{01}^{\prime }} &=&g_{1}\varphi
_{2}^{\prime }\varphi _{3}^{\prime }+(f_{1}\xi _{1}+f_{2}\xi _{2})\varphi
_{1}^{\prime }=0  \notag \\
\frac{\partial w}{\partial \varphi _{02}^{\prime }} &=&g_{1}\varphi
_{3}^{\prime }\varphi _{1}^{\prime }+(f_{1}\xi _{1}+f_{2}\xi _{2})\varphi
_{2}^{\prime }=0  \notag \\
\frac{\partial w}{\partial \varphi _{03}^{\prime }} &=&g_{1}\varphi
_{1}^{\prime }\varphi _{2}^{\prime }+(f_{1}\xi _{1}+f_{2}\xi _{2})\varphi
_{3}^{\prime }=0  \notag  \label{phiprime}
\end{eqnarray}

And also
\begin{equation}
\frac{\partial w}{\partial \xi_{0}}=f_{3}(\varphi^{\prime}\varphi^{%
\prime})+f_{ij}\xi{i}\xi{j}=0  \label{xi}
\end{equation}
which sets the magnitude of $\varphi^{\prime}\varphi^{\prime}$.

To be able to satisfy eqs.(\ref{phiprime}) while having the magnitude of $%
\varphi ^{\prime }$ fixed by eq(\ref{xi}), the VEVs of the singlets must be
such to make $f_{1}\xi _{1}+f_{2}\xi _{2}$ vanish. That leaves us with the
solution
\begin{equation}
\varphi ^{\prime }=(0,0,v^{\prime }),  \notag  \label{phipvev}
\end{equation}%
where at tree level $v^{\prime }$ is undetermined but will be
induced through dimensional transmutation at radiative order if
radiative corrections drive the $\varphi ^{\prime }$ mass squared
negative. Such radiative correction are generic and occur if the
field $\varphi ^{\prime }$ has significant Yukawa couplings such
as the $g_{1}$ term in Eqs.(\ref{wd}, \ref{phivev},\ref{phipvev})
generates the required vacuum alignment.

Note that the potential presented here has an important advantage
over the potential considered in \cite{Altarelli:2005yp} in that
the associated $A_{4} $ model does not require the vanishing of
any coupling allowed by the symmetry - at the cost of including
one extra singlet field. Such \textquotedblleft
supernatural\textquotedblright\ vanishing was necessary in the
supersymmetric model constructed by Altarelli and Feruglio and led
them to construct a five dimensional model in order to obtain a
fully natural theory. Our example here shows that this version of
the four dimensional model is also fully natural - the remainder
of the model is identical to that presented in
\cite{Altarelli:2005yp} - and leads to tri-bimaximal mixing. This
model has recently been constructed by Altarelli and Feruglio in a
paper we received while completing this work \cite{altlatest}.

\subsection{$Z_{3}^{\prime }\ltimes Z_{2}$}

The group $A_{4}$ has the structure of the semi-direct product group $%
Z_{3}^{\prime }\ltimes Z_{2}$ and its structure can help to understand the
properties of the group and the nature of the group invariants. Under it a
generic $A_{4}$ \textquotedblleft triplet\textquotedblright\ field $\phi _{i}
$ transforms in the manner given in Table \ref{so3discrete}. From this it is
clear that the only low order invariants are $\phi ^{2}=\phi _{i}$ $\phi _{i}
$ and $\phi ^{3}=\phi _{1}$ $\phi _{2}$ $\phi _{3}$ as used above. The $%
Z_{3}^{\prime }$ and $Z_{2}$ factors are clearly discrete subgroups of $SO(3)
$ and thus one sees that the $Z_{3}^{\prime }\ltimes Z_{2}$ non-Abelian
group is also a subgroup of $SO(3).$

Given this it is easy to generalise the $SO(3)$ model discussed above%
\footnote{%
This example demonstrates that the vacuum structure is natural but if one
wants to embed the symmetry breaking sector into the model of \cite%
{King:2005bj} it will be necessary to extend the additional
symmetry slightly.} by reducing the symmetry group to
$Z_{3}^{\prime }\ltimes Z_{2}$ and identifying the fields $\phi
_{123}=\varphi $ and $\phi _{3}=\varphi ^{\prime }.$ To get the
full model it is also necessary to generate $\phi
_{23}=(0,v^{\prime \prime },-v^{\prime \prime })$ for the
remaining flavon field. Its alignment is readily obtained.
Introduce the singlet driving fields $\chi _{0}$ and $\chi _{1}$
which transform as in Table \ref{x1}. The allowed superpotential
terms are
\begin{equation}
w^{\prime }=h_{1}\chi _{0}\phi _{23}\phi _{123}\phi _{123}+h_{2}\chi
_{1}\phi _{3}\phi _{23}\phi _{23}.  \label{wp}
\end{equation}

If radiative corrections drive the mass squared of the field $\phi _{23}$
negative at the scale $\Lambda $, it will acquire a VEV of $O(\Lambda )$.
The condition $F_{\chi _{0}}=0$ forces this vev to be orthogonal to that of $%
\phi _{123}$ and $F_{\chi _{1}}$ fixes its orientation relative to $\phi _{3}
$ giving the VEV
\begin{equation}
\phi _{23}=(0,-v^{\prime \prime },v^{\prime \prime }) , \text{ \ \ \ \ \ }%
v^{\prime \prime }\simeq \Lambda \text{\ }  \label{chip}
\end{equation}%
Now $\phi _{3}$, $\phi _{123}$ and $\phi _{23}$ generate tri-bimaximal
mixing using the strategy illustrated in section \ref{Sec:CSD} and developed
in \cite{King:2005bj}.

\qquad \qquad \qquad \qquad \qquad \qquad \qquad \qquad \qquad
\begin{table}[tbp] \centering%
\begin{tabular}{|l|l|l|}
\hline
$\phi _{i}$ & $Z_{3}^{\prime}\phi |_{i}$ & $Z_{2}\phi |_{i}$ \\ \hline
$\phi _{1}$ & $\rightarrow \phi _{2}$ & $\rightarrow \phi _{1}$ \\
$\phi _{2}$ & $\rightarrow \phi _{3}$ & $\rightarrow -\phi _{2}$ \\
$\phi _{3}$ & $\rightarrow \phi _{1}$ & $\rightarrow -\phi _{3}$ \\ \hline
\end{tabular}%
\caption{Transformation properties of a generic triplet field $\phi$ under the
semi direct product group $Z_{3}'\ltimes Z_{2}$.}\label{so3discrete}%
\end{table}%

\begin{center}
\begin{table}[tbp] \centering%
\begin{tabular}{|l|l|l|l|l|l|l|l|}
\hline \texttt{Field} & $\phi_{123}$ & $\phi_{3}$ & $\phi_{23} $ &
$\varphi _{0}$ & $\varphi _{0}^{\prime}$ & $\chi _{0}$ & $\chi
_{1}$ \\ \hline
$Z_{3}$ & $1$ & $\omega $ & $1$ & $1$ & $\omega $ & $1$ & $\omega^2$ \\
$U(1)_{R}$ & $0$ & $0$ & $1$ & $2$ & $2$ & $1$ & $0$ \\ \hline
\end{tabular}

\caption{Transformation property of the fields in the $Z_{3}'\ltimes Z_{2}$
model.} \label{x1}%
\end{table}%
\end{center}

\section{$Z_{3}^{\prime }\ltimes Z_{3}^{\prime \prime }\ltimes Z_{2}$ $%
(\Delta (108))$}

In the model based on $Z_{3}^{\prime }\ltimes Z_{2}$ the left-handed $%
SU(2)_{L\text{ }}$ doublet fermions, $\psi _{i}$, are triplets under the $%
Z_{3}^{\prime }$ while the left-handed charge conjugate $SU(2)_{L\text{ }}$
singlet fermions, $\psi _{i}^{c}$, are singlets. As a result it is not
straightforward to embed the model in an underlying $SO(10)$ theory. In this
Section we show how vacuum alignment through a non-Abelian discrete symmetry
can readily be consistent with an underlying $SO(10)$ structure.

As a simple example consider the discrete group $Z_{3}^{\prime }\ltimes
Z_{3}^{\prime \prime }\ltimes Z_{2}$ \footnote{%
This group is $\Delta (108)$, i.e. the dihedral like discrete subgroup of $%
SU(3)$ with $n=6$ \cite{Klink}} in which triplet fields $\phi
_{i\text{ }}$transform as shown in Table \ref{su3discrete} where
$\omega $ is the cube root of unity. In this case the only low
order invariant allowed by this symmetry is $\phi ^{3}=\phi _{1}$
$\phi _{2}$ $\phi _{3}$. The reason this is important is
because an underlying $SO(10)$ gauge group requires that $\psi _{i}$ and $%
\psi _{i}^{c}$ should be assigned to the same triplet representation. In
order to build a viable model of masses it is necessary to forbid the
invariant $\psi _{i}\psi _{i}^{c}$. This is possible with discrete subgroups
of $SU(3)$ family symmetry as this example shows (but is not possible for
discrete subgroups of $SO(3)$ family symmetry as the previous example
demonstrated).

Apart from this difference, the model is quite similar to the previous
example with fields $\phi _{3},$ $\phi _{23},$ $\phi _{123}$ as in the
previous example which transform under the same symmetry with the same
charges as in Table \ref{x1}. In this case the superpotential takes the form

\begin{eqnarray}
w &=& g(\varphi _{0}\phi_{123} \phi_{123} ) +g_{1}(\varphi
_{0}^{\prime}\phi_3 \phi_3) +\frac{h_{1}}{M^3}(\varphi _{0}\phi_{123}
\phi_{123}) (\phi_{123}\phi_{123}\phi_{123} )  \notag \\
&+& \frac{h_{2}}{M^3}(\varphi _{0,1}\phi_{123,1}^{5}+
\varphi_{0,2}\phi_{123,2}^{5}+\varphi_{0,3}\phi_{123,3}^{5})
\end{eqnarray}%
Here we have allowed for the two possible higher dimension terms of the form
$(\varphi \phi \phi )(\phi \phi \phi )$ and $(\varphi
_{1}\phi_{1}^{5}+\varphi _{2}\phi _{2}^{5} +\varphi _{3}\phi _{3}^{5})$
because, unlike the first example, the vacuum structure is sensitive to such
higher order terms in leading order. The scale $M$ is the messenger mass
scale, possibly the Planck scale $M_{Planck},$ responsible for generating
these operators.

Clearly the vacuum structure of $\phi _{3}$ is still determined by the Eq.(%
\ref{phiprime}) so, allowing for radiative breaking we have
\begin{equation}
\phi _{3}=(0,0,v^{\prime })~~~.~~~~~~~~~~~~~~  \notag
\end{equation}%
However the minimisation conditions for $\phi _{123}$ change and are now
given by
\begin{eqnarray}
\frac{\partial w}{\partial \varphi _{0,1}} &=&\Phi _{2}\Phi
_{3}(g+h_{1}(\Phi \Phi \Phi ))+h_{2}\Phi _{1}^{5}=0  \notag \\
\frac{\partial w}{\partial \varphi _{0,2}} &=&g\Phi _{3}\Phi
_{1}(g+h_{1}(\Phi \Phi \Phi ))+h_{2}\Phi _{2}^{5}=0  \notag \\
\frac{\partial w}{\partial \varphi _{0,3}} &=&g\Phi _{1}\Phi
_{2}(g+h_{1}(\Phi \Phi \Phi ))+h_{2}\Phi _{3}^{5}=0
\end{eqnarray}%
where on the right-hand side we have written $\Phi =\phi _{123}$. This is
solved by
\begin{equation}
\phi _{123}=(v,v,v),~~~~~~~v^{3}=-\frac{gM^{3}}{(h_{1}+h_{2})}~~~.
\end{equation}%
Once again we see, by suitable choice of parameters, that it is easy to
obtain the vacuum alignment needed for tri-bimaximal mixing. The vev of the
field $\phi _{23}$ in the direction given by Eq.(\ref{chip}) may be aligned
in the same way as the $Z_{3}^{\prime }\ltimes Z_{2}$ model thorough the
introduction of the singlet driving fields $\chi _{0}$ and $\chi _{1}$ which
transform as in Table \ref{x1}, giving the the allowed superpotential terms,
of Eq.(\ref{wp}). The full model based on this discrete symmetry subgroup of
$SU(3)$ is a simplification of the model given in \cite%
{deMedeirosVarzielas:2005ax}, and will be discussed in a future publication
\cite{future}.

\begin{table}[tbp] \centering%
\begin{tabular}{|l|l|l|l|}
\hline
$\phi _{i}$ & $Z_{3}^{\prime }\phi |_{i}$ & $Z_{3}^{\prime \prime }\phi |_{i}
$ & $Z_{2}\phi |_{i}$ \\ \hline
$\phi _{1}$ & $\rightarrow \phi _{2}$ & $\rightarrow \phi _{1}$ & $%
\rightarrow \phi _{1}$ \\
$\phi _{2}$ & $\rightarrow \phi _{3}$ & $\rightarrow \omega \phi _{2}$ & $%
\rightarrow -\phi _{2}$ \\
$\phi _{3}$ & $\rightarrow \phi _{1}$ & $\rightarrow \omega ^{2}\phi _{3}$ &
$\rightarrow -\phi _{3}$ \\ \hline
\end{tabular}%
\caption{Transformation properties of a generic triplet field $\phi$ under the
semi direct product group $Z_{3}'\ltimes Z_{3}^{\prime \prime }\ltimes
Z_{2}$.} \label{su3discrete}%
\end{table}%

In summary, tri-bimaximal mixing in the neutrino sector occurs
quite naturally in CSD models in which vacuum alignment follows
from a discrete non-Abelian subgroup of the $SU(3)$ maximal family
group commuting with an underlying GUT. In our examples the
tri-maximal mixing is directly related to the existence of the underlying $%
Z_{3}$ factor while the bi-maximal mixing is due to the $Z_{2}$
factor, giving a very intuitive origin for the structure. The
strategy we have detailed here allows for the extension of a Grand
Unified Theory to include a non-Abelian family symmetry of this
type. While it seems impossible to incorporate the full $SU(3)$
family symmetry in heterotic or D-brane string constructions, such
discrete non-Abelian groups readily appear as symmetries of the
underlying compactification manifold. This is encouraging for the
prospect of building a viable superstring theory of fermion masses

\section*{Acknowledgements}

The work of I. de Medeiros Varzielas was supported by FCT under the grant SFRH/BD/12218/2003.
This work was partially supported by the EC 6th Framework Programme MRTN-CT-2004-503369.

\end{document}